\documentclass{article}
\usepackage{epsfig}
\begin{document} 
\Large{Protostellar jets and magnetic diffusion}

\large
Miljenko \v{C}emelji\'{c}$^1$ and Christian Fendt$^{1,2}$
 
$^1$Astrophysikalisches Institut Potsdam (Germany), $^2$Universit\"at Potsdam,
 Institut f\"ur Physik; email: cemeljic,cfendt@aip.de
\begin{abstract}
We investigate the evolution of a disk wind into a collimated 
jet under the influence of magnetic diffusivity.
Using the ZEUS-3D code in the axisymmetry option we solve the
time-dependent resistive MHD equations for a model setup of a central
star surrounded by an accretion disk.
The disk is taken as a time-independent boundary condition for the
mass flow rate and the magnetic flux distribution.
We find that the diffusive jets propagate slower into the ambient
medium, most probably due to the lower mass flow rate in axial
direction.
Close to the star we find that a quasi stationary state evolves after
several hundreds (weak diffusion) or thousands (strong diffusion)
of disk rotations.
Magnetic diffusivity affects the protostellar jet structure de-collimating
it. We explain these effects in the framework of the Lorentz force.
\end{abstract}
keywords: accretion, accretion disks --
           MHD --
           ISM: jets and outflows --
           stars: mass loss --
           stars: pre-main sequence              

\section{Magnetic jets from accretion disks}  

Astrophysical jets detected so far seem to be attached to objects where an
accretion disk is indicated to be present. The similarities between jets 
from the different sources imply that the basic mechanism for jet formation 
should be the same.

\begin{figure} 
\parbox{40mm}
\thicklines
\epsfysize=40mm
\put(0,0){\epsffile{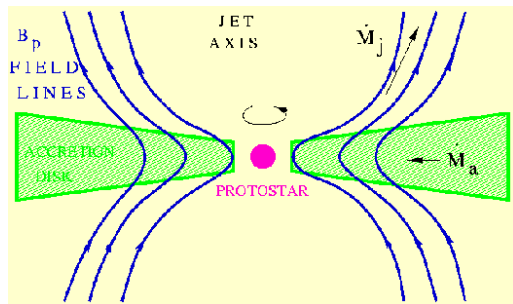}}
\caption[]{Sketch of the model.}
\label{f1}
\end{figure}

What kind of mechanism turns the inflowing matter of the accretion disk into
an outflow from the disk (or the star) is still not really known, although
it seems to be clear that magnetic fields play a major role.

Our model is sketched in Figure~\ref{f1}. In order to model the time-dependent 
evolution of jet formation, we solve the set of resistive MHD equations,

\vspace{1cm}

\begin{equation}
{\frac{\partial \rho}{\partial t}} + \nabla \cdot (\rho \vec{v} ) = 0
\end{equation}

\begin{equation}
\rho \left[ {\frac{\partial\vec{u} }{\partial t}}
+ \left(\vec{v} \cdot \nabla\right)\vec{v} \right]
+ \nabla (p+p_A) +
 \rho\nabla\Phi - \frac{\vec{j} \times \vec{B}}{c} = 0
\end{equation}

\begin{equation}
{\frac{\partial\vec{B} }{\partial t}}
- \nabla \times \left(\vec{v} \times \vec{B}
-{\frac{4\pi}{c}} \eta\vec{j}\right)= 0
\end{equation}

\begin{equation}
e=p/(\gamma-1)
\end{equation}

\begin{equation}
\nabla \cdot\vec{B} = 0
\end{equation}

\begin{equation}
\frac{4\pi}{c} \vec{j} = \nabla \times \vec{B}
\ .
\end{equation}

\vspace{1cm}

We apply a polytropic equation of state, $p=K\rho^\gamma$ with a polytropic index $\gamma=5/3$. We do not solve the energy  equation but define the internal energy of 
the system with eq. (4). We may expect that the turbulence pattern in the disk may also enter the disk corona and the jet, and that the jet flow itself is subject to turbulent diffusion. The magnetic diffusivity is denoted by the variable $\eta$, and other have its 
usual meaning, with $p_{\rm A}\equiv p/\beta_{\rm T}$ with $\beta_{\rm T} = {\rm const.}$, an Alfv\'{e}nic turbulent pressure, added to the hydrostatic pressure $p$.

\section{Results and discussion}
\begin{figure} 
\parbox{120mm}
\thicklines
\epsfysize=40mm
\put(0,250){\epsffile{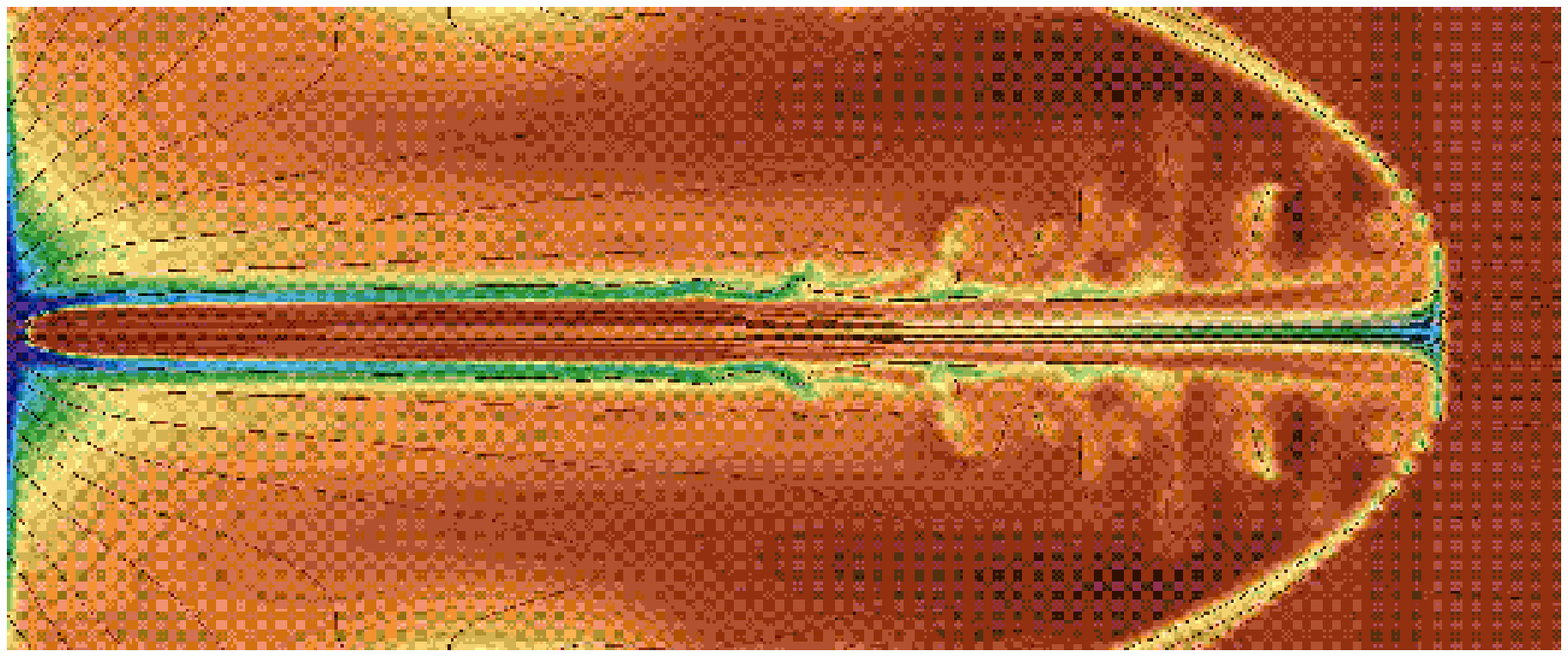}}
\put(0,125){\epsffile{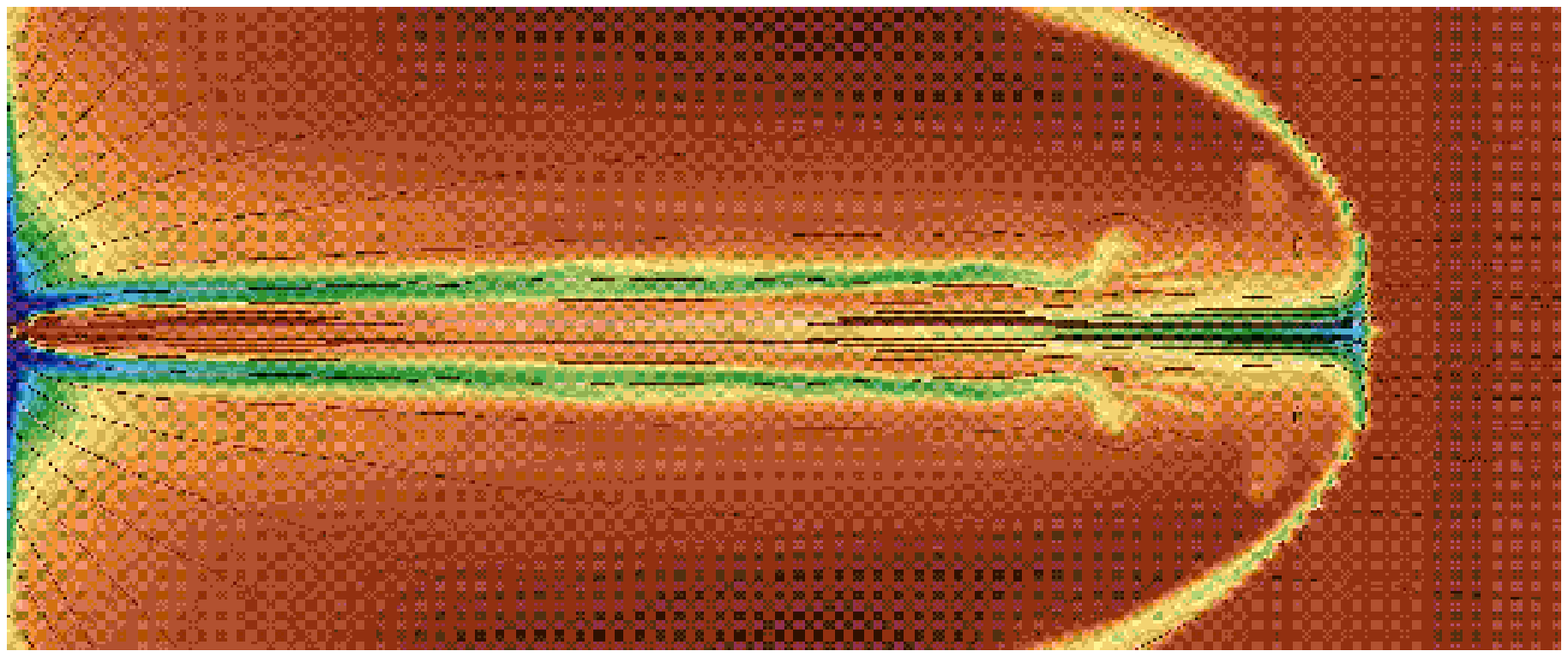}}
\put(0,0){\epsffile{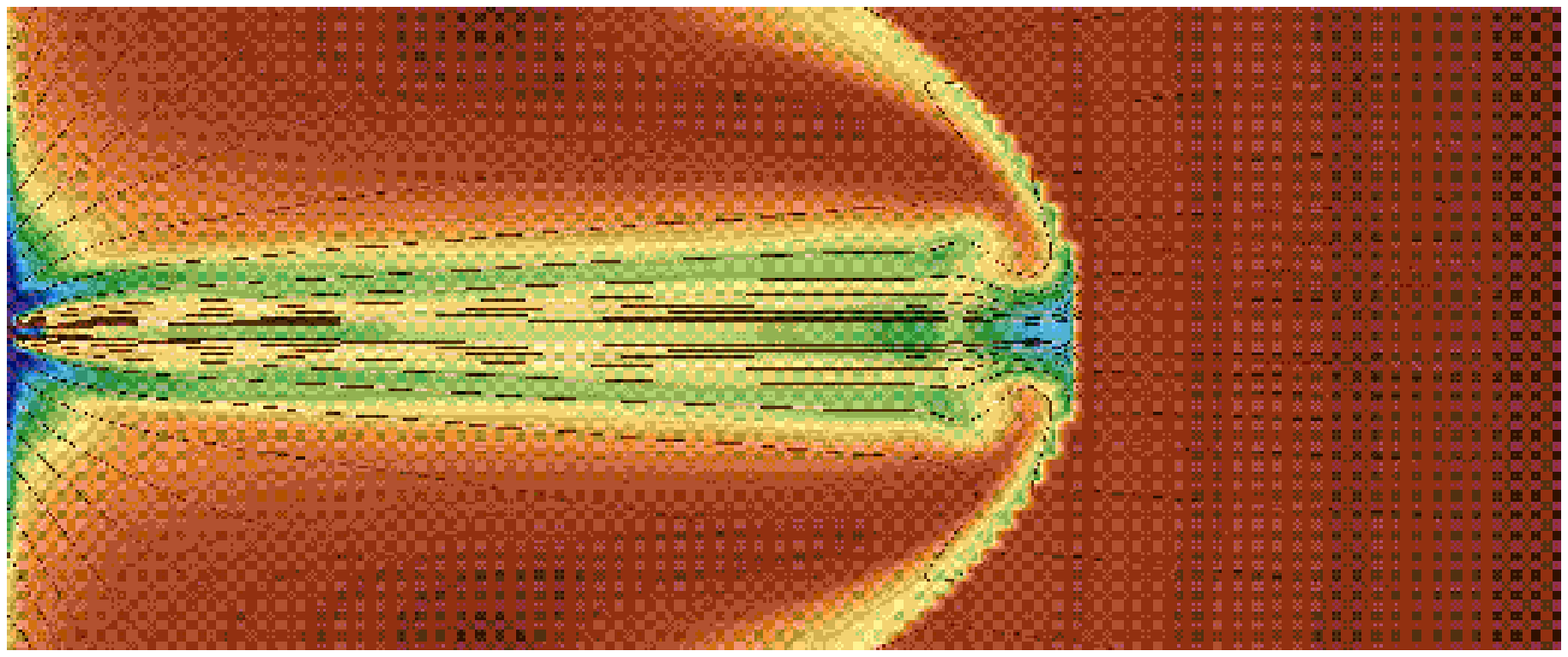}}
\caption[]{Global evolution of the jet on a grid of 
(z$\times$r)=(280$\times$40)r$_{\rm i}$ with a resolution 
of 900$\times$200 elements. Shown is the state of evolution after t=400 rotations of the disk inner
radius for different magnetic diffusivity, $\eta$=0,0.01,0.1 for top, 
middle and bottom panel, respectively.
Colors (blue to yellow in decreasing manner) indicate density.
Lines denote twenty linearly spaced magnetic flux
surfaces (or poloidal field lines).
The figure demonstrates that the bow shock advances slower with increasing
diffusivity.}
\label{f2}
\end{figure}

The results of one reference set of global simulations with high resolution
(numerical mesh of $900\times 200$ grid points,
physical grid of  $(z\times r)=(280\times 40)r_{\rm i}$) are presented 
in Figure~\ref{f2}.

\vspace{1cm}

\begin{figure} 
\vspace{-9pc}
\parbox{50mm}
\thicklines
\epsfysize=50mm
\put(0,0){\epsffile{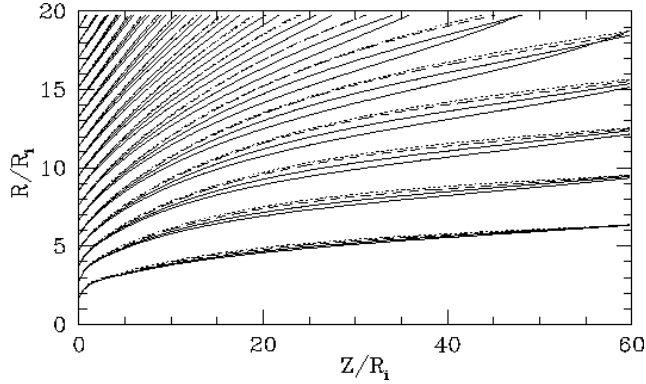}}
\caption[]{The evolution of the inner jet approaches the 
quasi-stationary state. Shown are poloidal magnetic field lines in the case of $\eta$ = 0.1
for different time steps, t = 250, 300, 350, 400
(thick solid, thin solid, dashed and dotted lines, respectively).
In these simulations beta plasma is half the value from the global simulations 
above, but this just changes the position of the Alfv\'{e}n surface above the 
disk slightly. Grid size is 280$\times$80 elements for a physical size of 
(140$\times$40)r$_{\rm i}$.
The picture shows how the poloidal magnetic field lines diffuse outwards
but approach a (quasi)-stationary state after 400 rotations (see the dashed 
and dotted lines almost coinciding).
}
\label{f3}
\end{figure}

In order to investigate only the gross behavior of the 
jet flow and not its structure in detail,
we run another set of simulations with lower resolution 
(numerical mesh of 280$\times$80 grid points, 
physical grid of  (z$\times$r)=(140$\times$40)r$_{\rm i}$) -- see Figure~\ref{f3}.

\vspace{1cm}

\begin{figure} 
\parbox{60mm}
\thicklines
\epsfysize=50mm
\put(190,0){\epsffile{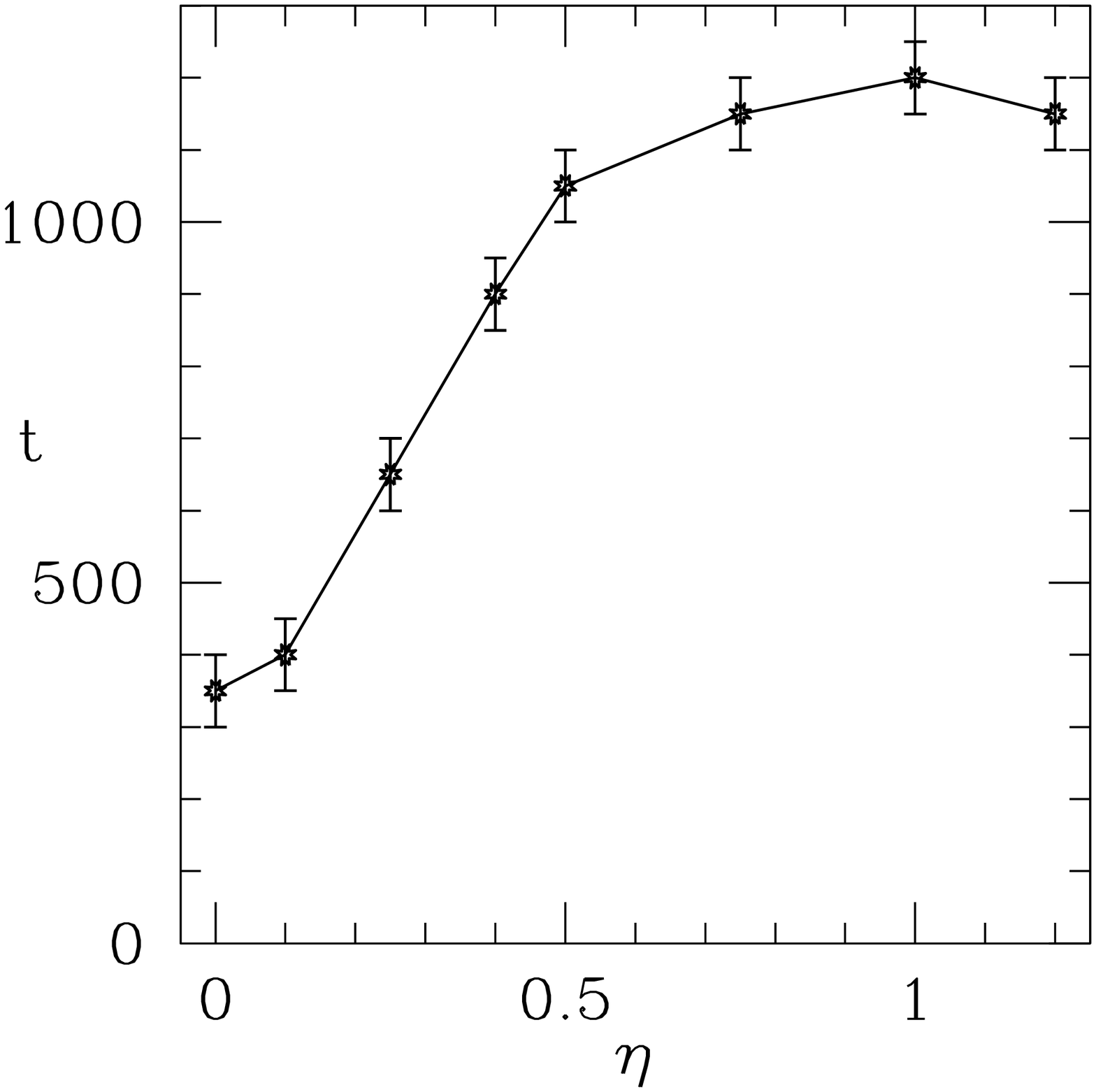}}
\epsfysize=45mm
\put(0,0){\epsffile{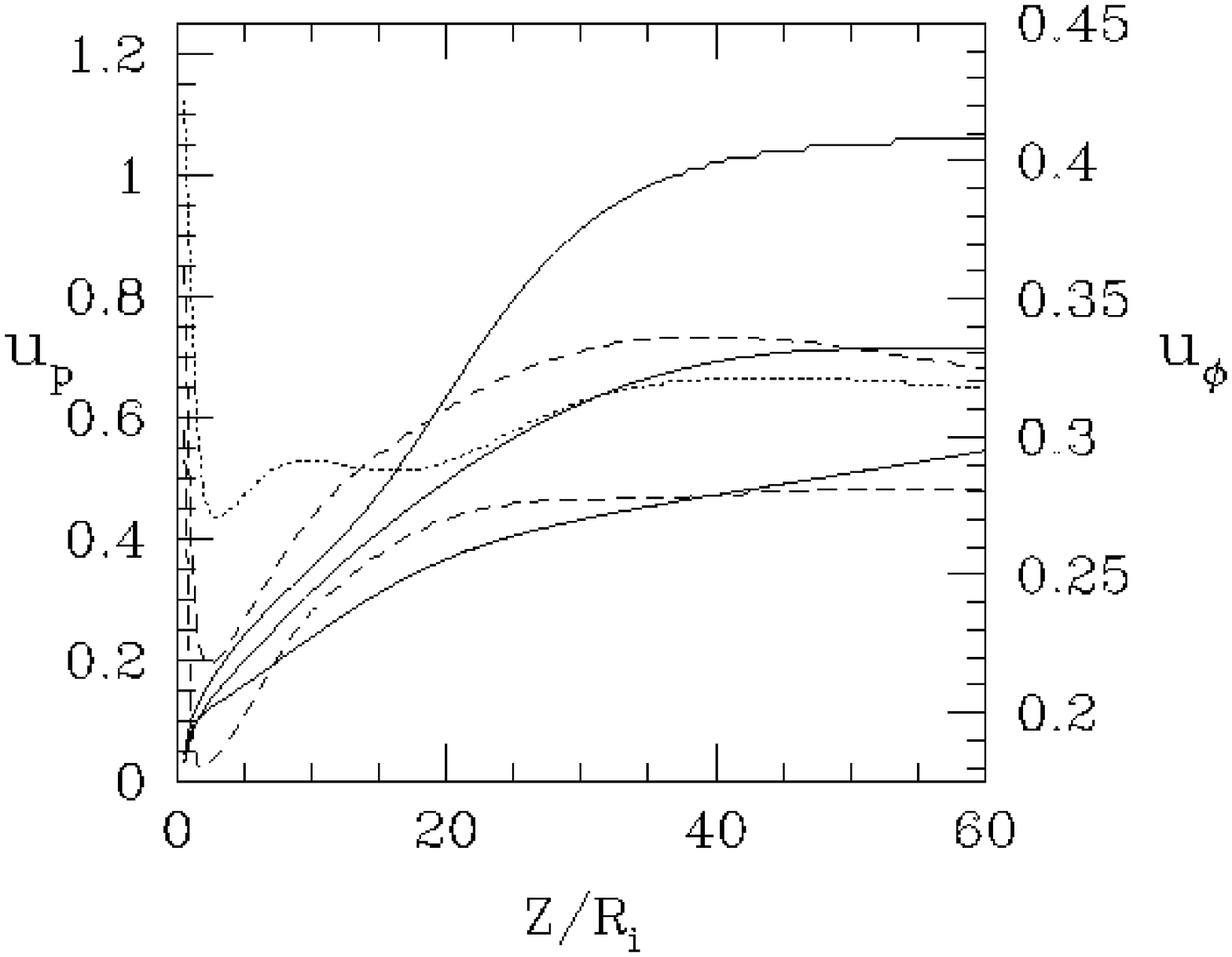}}
\caption[]{Left: Components of the velocities. Cuts in Z direction at 
r=15 for 
the quasistationary time of simulations. 
Poloidal components are given in solid lines, toroidal in dashed lines. Two 
different 
ordinate scales are used to expose relative changes of the velocities.  Left 
ordinate gives the values of the poloidal velocity, when right gives the values 
of the toroidal velocity. Both components increase with the 
increasing 
diffusivity. 
Right panel:  The times when the quasi-stationary state is reached, 
in dependence on diffusivity.
}
\label{f4}
\end{figure}

In apparent contrast with the bow shock propagation from Figure~\ref{f2} is the 
increase of the jet velocity with diffusivity (Figure~\ref{f4}, left panel).

Below the `'critical $\eta$`' in our choice of parameters ($\eta<0.5$), the 
time of reaching the quasi-stationary 
state of the inner jet is approximately linearly dependent on $\eta$ (Figure~\ref{f4}, right panel).

\begin{figure} 
\vspace{0pc}
\parbox{120mm}
\thicklines
\epsfysize=60mm
\put(0,0){\epsffile{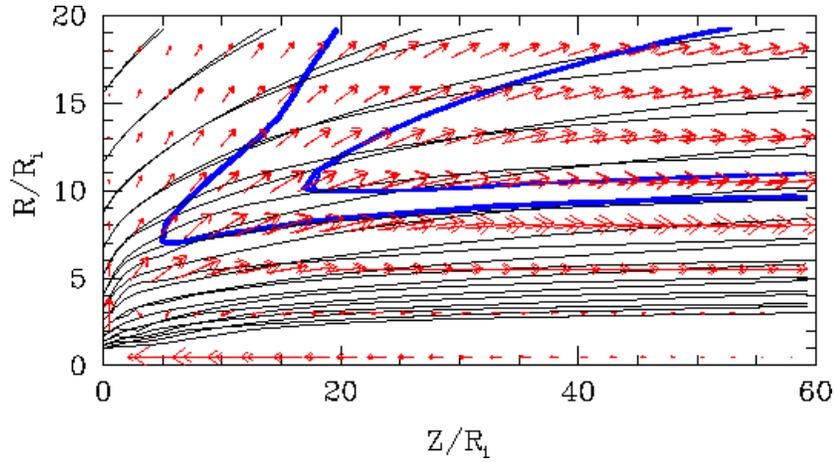}}
\caption[]{The de-collimation of the stationary state poloidal magnetic 
field due to magnetic diffusivity. Thick black lines for $\eta$=0, and thin 
lines for
$\eta$=0.1. Overly of the poloidal velocity vectors show that the velocity 
pattern is not de-collimated. Thick blue line is the Alfv\'{e}n 
surface, and thin blue line is the fast magnetosonic surface for the diffusive 
simulation.}
\label{f5}
\end{figure}

Interesting is to note that de-collimation of the poloidal magnetic field 
lines is 
not followed by the poloidal velocity vectors -- see Figure~\ref{f5}. 

The much faster computation of the low resolution simulations allowed us to
follow the jet evolution for a very long time 
even in the case of a high magnetic diffusivity (up to 4000 disk rotations).
The observed de-collimation of the matter flow with increasing
diffusivity is most evident if we plot the mass and momentum
fluxes across the boundaries of the inner jet region.

\begin{figure} 
\vspace{-2pc}
\parbox{70mm}
\thicklines
\epsfysize=70mm
\put(0,0){\epsffile{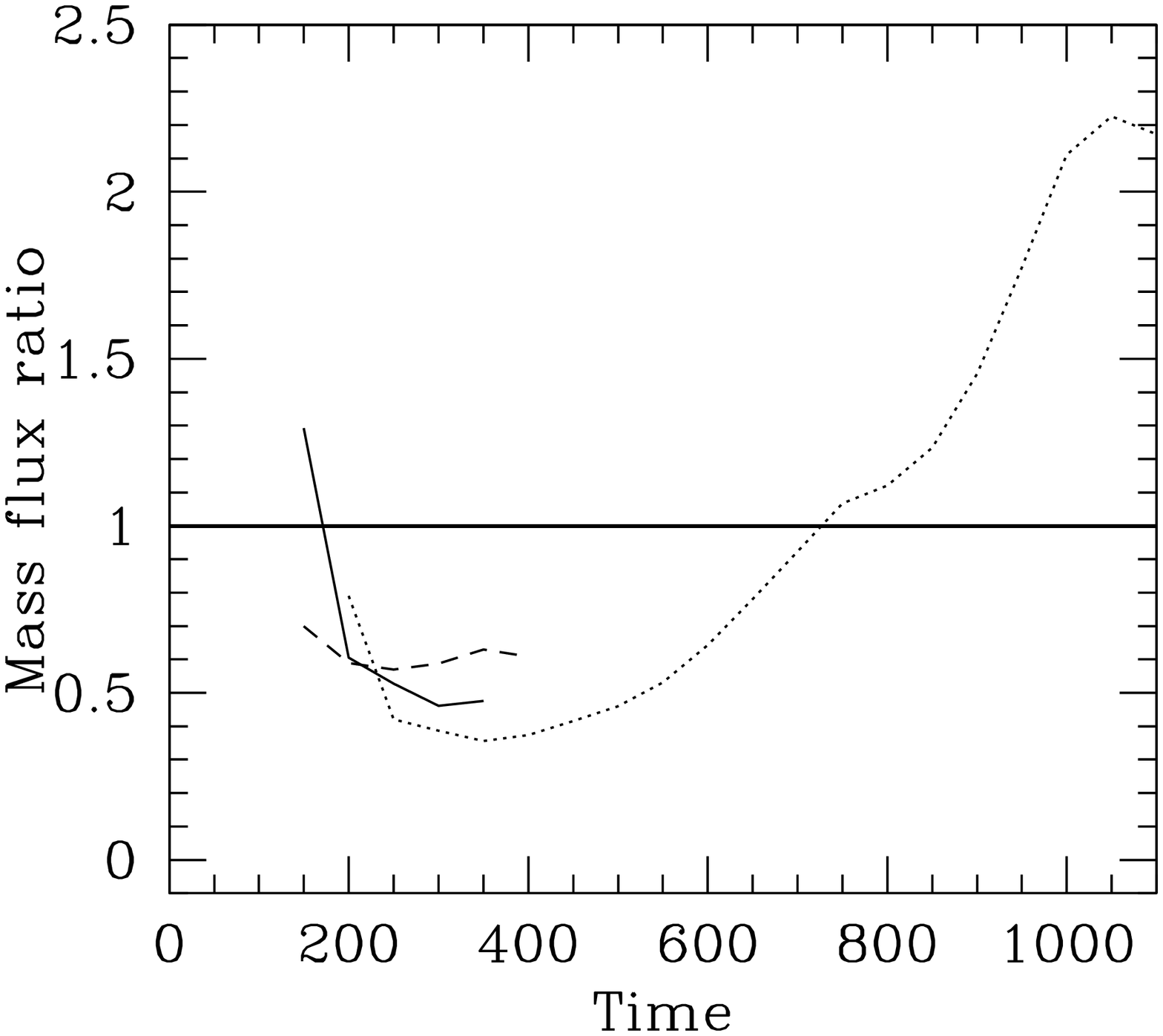}}
\caption[]{Time evolution of the mass flow ratio between the radial 
outflow boundary and the axial outflow boundary in the inner part of the jet, 
(z$\times$r)=(60$\times$20)r$_{\rm i}$. Parameter $\eta$=0,0.1,0.5 
in solid, dashed and dotted line, respectively. For higher diffusivity, the 
mass
 flux ratio in the quasi-stationary state increases, indicating a decrease in 
 degree of collimation.}
\label{f6}
\end{figure}

In combination with various physical effects --
magnetic and inertial forces, pressure and gravity --
the magnetic diffusion will modify the MHD structure of 
the jet.
As the flow approaches the Alfv\'en surface, a toroidal magnetic field
component is induced (`'wound-up`') due to the inertial back-reaction
of the matter on the field.
The toroidal field may lead to (de-) accelerating Lorentz forces 
$\vec{F}_{\rm L, ||} \sim \vec{j}_{\perp} \times \vec{B}_{\phi} $
and (de-) collimating forces
$\vec{F}_{L,\perp} \sim \vec{j}_{||} \times \vec{B} $ 
where, here, the perpendicular and parallel projection is made with
respect to the poloidal magnetic field 
(which, only in the case of ideal stationary MHD is parallel to the
poloidal velocity).

\begin{figure} 
\vspace{-1pc}
\parbox{60mm}
\thicklines
\epsfysize=60mm
\put(170,0){\epsffile{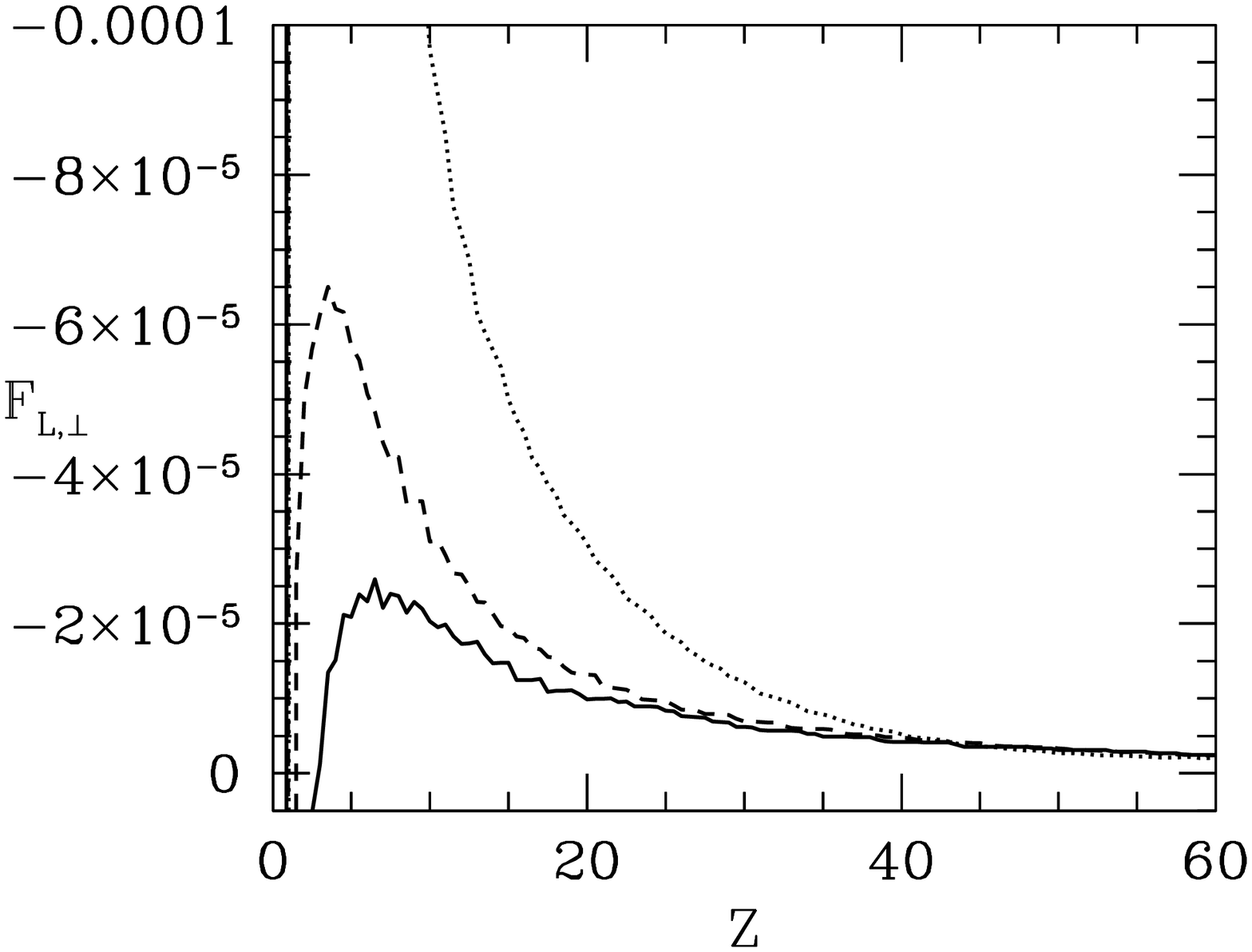}}
\epsfysize=60mm
\put(0,0){\epsffile{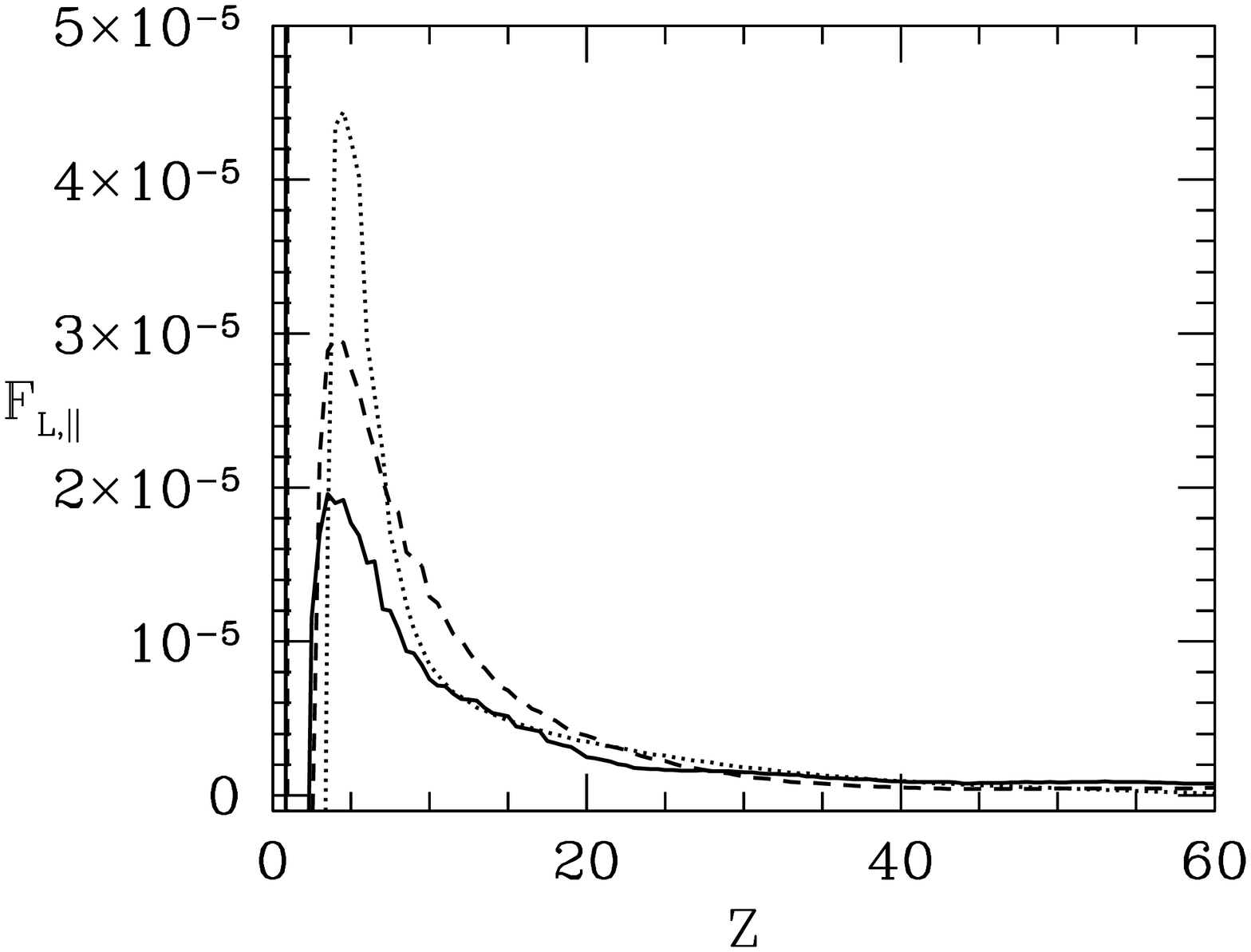}}
\vspace{-3pc}
\caption[]{Lorentz forces in the jet for different magnetic 
diffusivity
$\eta$ = 0, 0.1, 0.5 (solid, dashed and dotted lines, respectively).
 (Normalized) values of the force component
perpendicular (left panel) and parallel (right panel) 
to the field line. The toroidal component is similar to the parallel one. 
Components are projected
along a flux surface leaving the box of the inner jet close to 
(R=20,Z=60)-corner. The sign is defined as follows. 
For the perpendicular component the positive sign denotes the r-direction
(de-collimating force).
For the parallel component the positive sign denotes the z-direction 
(accelerating force).}
\label{f7}
\end{figure}
The winding-up of the poloidal magnetic fields is less efficient since 
magnetic diffusion leads to a slip of matter across the field.
The induced toroidal magnetic field is weaker leading to
a less efficient acceleration by Lorentz forces but also to a de-collimation.
As a de-collimation of the poloidal magnetic structure also implies a 
smaller launching angle for the sub-Alfv\'enic flow, the magneto-centrifugal
acceleration mechanism may work more effective.

acknowledgements:
We thank the LCA team and M.~Norman for the possibility to use the
ZEUS-3D code. This work was partly financed by the  DFG Schwerpunktprogramm 
``Physik der Sternentstehung''
(FE490/2-1).

\end{document}